\documentclass[aps,prl,twocolumn,groupedaddress]{revtex4}

\usepackage{graphicx}

\begin{document}


\title{Self-consistent crystalline condensate in chiral Gross-Neveu and Bogoliubov-de Gennes systems}


\author{G\"ok\c ce Ba\c sar and Gerald V.~Dunne}
\affiliation{Physics Department, University of Connecticut, Storrs CT 06269}



\begin{abstract}
We derive a new exact self-consistent crystalline condensate in the $1+1$ dimensional chiral Gross-Neveu model. This also yields a new exact crystalline solution for the one dimensional Bogoliubov-de Gennes equations and the Eilenberger equation of semiclassical superconductivity. We show that the functional gap equation can be reduced to a solvable nonlinear equation, and discuss implications for the temperature-chemical potential phase diagram.

\end{abstract}


\maketitle


Interacting fermion systems describe  a wide range of physical phenomena, from particle physics, to  solid state and atomic physics \cite{campbell,rajagopal,casalbuoni,pitaevskii}. Important paradigms include the Peierls-Frohlich model of conduction \cite{peierls}, the Gorkov-Bogoliubov-de Gennes approach to superconductivity \cite{degennes}, and the Nambu-Jona Lasinio (NJL) model of symmetry breaking in particle physics \cite{nambu}.  A $1+1$-dimensional version of the NJL model, the ${\rm NJL}_2$ model [also known as the {\it chiral} Gross-Neveu model, $\chi{\rm GN}_2$] has been widely studied as it exhibits asymptotic freedom, dynamical mass generation, and chiral symmetry breaking \cite{gross,dhn, shei,feinberg}. Surprisingly, the temperature-density phase diagram of this system is not yet  fully understood.  A gap equation analysis based on a homogeneous condensate suggests its phase diagram is the same as its discrete-chiral cousin, the original Gross-Neveu (${\rm GN}_2$) model \cite{gross}, while recent work finds an inhomogeneous Larkin-Ovchinikov-Fulde-Ferrell (LOFF) helical complex condensate (``chiral spiral'') below a critical temperature \cite{thies-njl}. In fact, the phase diagram of the discrete-chiral version, the ${\rm GN}_2$ model, has only recently been solved in the particle physics literature, analytically  \cite{thies-gn}, and on the  lattice  \cite{deforcrand}. There is a crystalline phase at low temperature and high density, and this phase is characterized by  a periodically inhomogeneous (real) condensate that solves exactly the gap equation.  This phase is not seen in the old phase diagram which was based on a uniform condensate \cite{wolff}. Interestingly, this discrete-chiral  ${\rm GN}_2$ model (with vanishing bare fermion mass) is mathematically equivalent to several models in condensed matter physics: the real periodic condensate may be identified with a polaron crystal in conducting polymers \cite{horovitz,braz,campbell}, with a periodic pair potential in quasi 1D superconductors \cite{mertsching,buzdin1}, and with the real order parameter for superconductors in a ferromagnetic field \cite{machida}. This system also is a paradigm of the phenomenon of fermion number fractionalization \cite{jackiw,gw}. Variants of such models also apply to ultracold fermionic systems, for which there are interesting new theoretical and experimental developments \cite{pitaevskii,zwierlein,partridge}. 

In this paper we present an analogous {\it complex} crystalline condensate for the {\it chiral} GN system, the ${\rm NJL}_2$ model. This condensate is an exact inhomogeneous solution to the gap equation, and also provides a new self-consistent solution to the Bogoliubov-de Gennes (BdG) \cite{degennes} and Eilenberger \cite{eilenberger} equations of superconductivity.  Our solution may also be relevant for chiral superconductors and for incommensurate charge density waves in quasi 1D systems \cite{sakita}, which have chiral symmetry and an inherently complex order parameter.

Consider the  massless ${\rm NJL}_2$ model with Lagrangian 
\begin{equation}
{\mathcal L}=\bar{\psi}\,i\, \partial \hskip -6pt / \,\psi +\frac{g^2}{2}\left[\left(\bar{\psi}\psi\right)^2+\left(\bar{\psi}i\gamma^5 \psi\right)^2\right]\quad ,
\label{lag}
\end{equation}
which has a continuous chiral symmetry $\psi\to e^{i\gamma^5 \alpha} \psi$. We have suppressed summation over $N$ flavors, which makes the semiclassical gap equation analysis exact in the $N\to\infty$ limit, a limit in which we can consistently discuss chiral symmetry breaking in 2D.
The original ${\rm GN}_2$ model \cite{gross}, without the pseudoscalar interaction term $\left(\bar{\psi}i\gamma^5 \psi\right)^2$, has a discrete chiral symmetry $\psi\to \gamma^5\psi$. 
There are two equivalent ways to find self-consistent static condensates. First, introduce bosonic condensate fields, $S\equiv \bar{\psi}\psi$ and $P\equiv \bar{\psi}i\gamma^5 \psi$, which we combine into a complex condensate field: $\Delta\equiv S-i\,P\equiv M\, e^{i\chi}$. Integrating out the fermion fields we obtain an effective action for the condensate as
\begin{equation}
S_{\rm eff}=-\frac{1}{2N g^2}\int |\Delta|^2 -i\,\ln\det\left[i\, \partial \hskip -6pt / \,-M\, e^{-i\gamma^5 \chi}\right]
\label{effective}
\end{equation}
The corresponding (complex) gap equation is
\begin{equation}
\hskip -2pt \Delta(x)=-2i N g^2 \frac{\delta}{\delta \Delta(x)^*}\ln\det\left[i\, \partial \hskip -6pt / \,-M(x)\, e^{-i\gamma^5 \chi(x)}\right]
\label{gap}
\end{equation}
One of the main results of this paper is that this gap equation can be reduced in an elementary manner to an ordinary differential equation, which moreover is soluble. A second approach to finding a self-consistent condensate is to solve the relativistic Hartree-Fock problem $H\psi=E\psi$, subject to the consistency condition $\langle\bar{\psi}\psi\rangle-i \langle\bar{\psi} i \gamma^5\psi\rangle=-\Delta/g^2$, with single-particle Hamiltonian
\begin{equation}
H=-i\gamma^5 \frac{d}{dx}+\gamma^0 \,M(x)\, e^{-i\gamma^5 \chi(x)}
=
\begin{pmatrix}
{-i\frac{d}{dx}&\Delta(x)\cr \Delta^*(x) & i\frac{d}{dx}}
\end{pmatrix}
\label{ham}
\end{equation}
This is the Bogoliubov-de Gennes (BdG), or Andreev, hamiltonian, with $\Delta(x)$ playing the role of the order parameter \cite{degennes}. We have chosen Dirac matrices  $\gamma^0=\sigma_1$, $\gamma^1=-i\sigma_2$, and $\gamma^5=\sigma_3$, to emphasize the natural complex combination $\Delta=S-iP$. The key object in our analysis is the Gork'ov Green's function, and 
in particular 
its coincident-point limit, the ``diagonal resolvent'':
\begin{equation}
R(x; E)\equiv \langle x| \frac{1}{H-E} |x\rangle
\label{res}
\end{equation}
This is a $2\times 2$ matrix, and the spectral function is $\rho(E)=\frac{1}{\pi}{\rm Im}\,{\rm Tr}_{D,x}\,R(x; E+i\epsilon)$, where the trace is both a Dirac and spatial trace. Approximation methods, such as the gradient and semiclassical expansions,  of $R(x; E)$ have been widely studied \cite{stone}. In one spatial dimension, $R(x; E)$ can be written in terms of two independent solutions $\psi_{1,2}$ to the Dirac/BdG equation $H\psi=E\psi$. That is, $R(x; E)=\left(\psi_1\psi_2^T+\psi_2\psi_1^T\right)\sigma_1/(2W)$, with Wronskian $W=-i\psi_1^T\sigma_2\psi_2$. It follows immediately that $R(x; E)$ satisfies the following first order equation:
\begin{equation}
R^{\,\prime} \gamma^5=i\,[\gamma^5 (E-\gamma^0 M\, e^{-i\gamma^5 \chi}), R \gamma^5]
\label{dikii}
\end{equation}
In superconductivity, (\ref{dikii}) is known as the Eilenberger equation \cite{eilenberger}, and in mathematical physics as the Dik'ii equation \cite{dikii}. It is also straightforward to show that $R=R^\dagger$, $\det R=1/4$, and ${\rm tr}(R \,\gamma^5)=0$. 

Our main observation is that the gap equation (\ref{gap}) motivates a self-consistent ansatz form of the $2\times 2$ matrix $R(x; E)$, and when this is combined with the identity (\ref{dikii}), the exact self-consistent condensate and associated resolvent are completely  determined. To see this, note that the gap equation (\ref{gap}) can be viewed in two ways. First, write the log det term as
\begin{eqnarray}
\int_{-\infty}^\infty dE\,\rho(E)\frac{1}{\beta} \ln\left(1+e^{-\beta(E-\mu)}\right)
\label{logform}
\end{eqnarray}
All dependence on $\Delta(x)$ resides in the spectral function $\rho(E)$, via ${\rm Tr}\, R$, and so the simplest solution to the gap equation (\ref{gap}) is for  {\it diagonal} entries of the $2\times 2$ matrix $R$ [recall they are equal since ${\rm tr}(R \,\gamma^5)=0$] to be linear in $|\Delta|^2$. On the other hand, we can also express the gap equation, by performing the functional derivatives, as
\begin{eqnarray}
\Delta(x)=iNg^2\,{\rm tr}_{D,E}\left[\left(\gamma^0+\gamma^1\right)R(x; E)\right]
\label{gap2}
\end{eqnarray}
This suggests that  the {\it off-diagonal} entries of $R$ be proportional to $\Delta$ and $\Delta^*$. In fact, consistency between (\ref{dikii}) and (\ref{gap2}) introduces derivative terms, leading to 
\begin{equation}
R(x; E)={\mathcal N}(E)
\begin{pmatrix} {
a(E)+|\Delta |^2& b(E) \Delta-i \Delta^\prime \cr
b(E) \Delta^*+i \Delta^{\prime\,*} & a(E)+|\Delta |^2 }
\end{pmatrix}
\label{ansatz}
\end{equation}
The gap equation is  satisfied since we find ${\rm tr}_E {\mathcal N}(E)=0$.
With this  ansatz, the diagonal entries of (\ref{dikii}) are identically satisfied, while
the off-diagonal entries imply that the condensate $\Delta$ satisfy the equation
\begin{equation}
\Delta^{\prime\prime} -2|\Delta |^2\,\Delta +i\left(b-2 E\right)\Delta^\prime -2\left(a-E b\right)\Delta=0
\label{nlse}
\end{equation}
This nonlinear Schr\"odinger equation (NLSE) is analytically soluble, 
and all previously known examples of self-consistent condensates in ${\rm GN}_2$ and ${\rm NJL}_2$ are special cases. Furthermore, the corresponding Dirac/BdG equation $H\psi=E\psi$ is also exactly soluble, and has a  spectrum consisting of a {\it single band} 
in the gap \cite{bd}. This provides an elementary explanation of the result from inverse scattering \cite{dhn,shei,feinberg} that self-consistent ground state condensates are reflectionless (or finite-gap) with a single state (or band).
For example, if we specialize to a real condensate, as is relevant for the [discrete-chiral] ${\rm GN}_2$ model, the [rescaled] self-consistent  solution is $\Delta(x)=\sqrt{\nu}\,{\rm sn}(x;\nu)$, which satisfies $\Delta^{\prime\prime}-2\Delta^3=-(1+\nu)\Delta$, where sn is a Jacobi elliptic function, with elliptic parameter $0\leq \nu\leq 1$ \cite{horovitz,braz,mertsching,buzdin1,machida}. The fermion spectrum has a single band in the gap, centered on $E=0$, reflecting the charge-conjugation symmetry of the ${\rm GN}_2$ system.
As $\nu\to1$, the period becomes infinite and we obtain the famous kink, $\Delta(x)={\rm tanh}(x)$, with a single bound state at $E=0$. This mid-gap zero mode has many interesting physical consequences in polymer systems, and is the paradigm of the fractional fermion number phenomenon \cite{jackiw}. It is also worth noting that for the [discrete-chiral] ${\rm GN}_2$ model with a bare fermion mass, the NLSE (\ref{nlse}) acquires an inhomogeneous term, in which case the general solution is written as $\Delta(x)=\zeta(\gamma)+\zeta(x)-\zeta(x+\gamma)$, where $\zeta$ is the Weierstrass zeta function, and this represents a kink-antikink crystal (or bipolaron crystal in the polymer language \cite{braz2}), which is a periodic generalization \cite{thies-gn} of the Dashen-Hasslacher-Neveu (DHN) kink-antikink solution \cite{dhn}: $\Delta(x)=\coth(b)+\tanh(x)-\tanh(x+b)$.

\begin{figure}[h]
\includegraphics[scale=0.8]{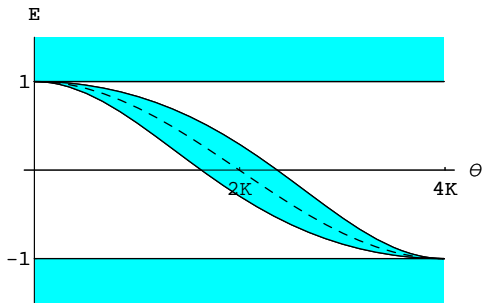}
\caption{Band spectrum of the BdG Hamiltonian (\ref{ham}) for the complex crystal condensate (\ref{complex}), as a function of the twist parameter $\theta$. The dashed line denotes the infinite period limit [in this case $\nu\to 0$], which is Shei's solution (\ref{shei}). At $\theta=2{\bf K}$ we recover the symmetric spectrum of the real kink crystal, relevant to the ${\rm GN}_2$ system.}
\label{sheispectrum}
\end{figure}

All these are {\it real} condensates, and are well-known. The only previously known complex  condensates are (i) the simple "chiral spiral" or LOFF solution $\Delta(x)=A e^{i q x}$; and (ii) Shei's "twisted kink" solution \cite{shei}, which we can express in complex form 
\begin{equation}
\Delta(x)=e^{i\theta/2}\,\frac{\cosh\left(x\,\sin \frac{\theta}{2}-i\frac{\theta}{2}\right)}{\cosh\left(x\,\sin \frac{\theta}{2}\right)}\quad .
\label{shei}
\end{equation}
$\theta$ is the angle through which the phase of the condensate rotates as $x$ goes from $-\infty$ to $+\infty$. The single-particle fermion spectrum has a single bound state within the gap, located at  $E=\cos\left(\frac{\theta}{2}\right)$, as shown by the dashed line in Figure \ref{sheispectrum}. 
Consistency with the gap equation requires vanishing of the coefficient of $\Delta^\prime$, which places conditions on parameters of the solution. For example, for Shei's solution the condition is that  $\theta/(2\pi)$ is equal to the filling fraction of the bound level \cite{shei,feinberg}. 
We point out a simple physical interpretation of this condition in terms of conserved currents (see also \cite{karbstein}). Both  ${\rm NJL}_2$ and ${\rm GN}_2$ models have a conserved current $j^\mu=\bar{\psi}\gamma^\mu\psi$. Since $\langle j^\mu(x)\rangle=-iN{\rm tr}_{D,E}[\gamma^0\gamma^\mu R(x)]$, we see that, for a static condensate, current conservation follows trivially since $\langle j^1(x)\rangle=0$; this is  the physical origin of the identity ${\rm tr}_{\rm D}\left(\gamma^5 R\right)=0$. The ${\rm NJL}_2$ model also has a conserved axial current, $j^\mu_5=\bar{\psi}\gamma^\mu\gamma^5\psi$, and the Eilenberger equation  (\ref{dikii}) implies 
\begin{eqnarray}
\partial_\mu \langle j^\mu_5\rangle\equiv -iN{\rm tr}(R^\prime)=2S(x)\langle \bar{\psi}\, i \gamma^5\psi\rangle -2 P(x)\langle \bar{\psi}\psi\rangle\,\,\,
\label{axial}
\end{eqnarray}
Axial current conservation then follows from the gap equation, using precisely the same condition on the coefficient of $\Delta^\prime$. 

\begin{figure}[h]
\includegraphics[scale=0.5]{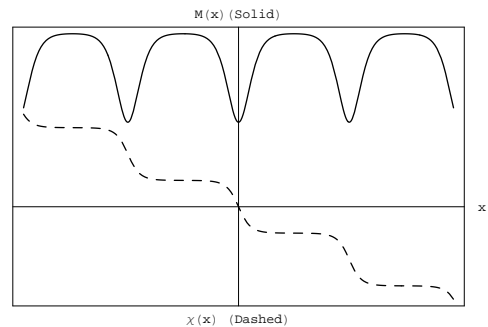}
\caption{The amplitude $M(x)$ [solid line] and phase $\chi(x)$ [dashed line] of $\Delta=M e^{i\chi}$ in (\ref{complex}), over several periods. The amplitude is periodic while the phase rotates by an angle $2\varphi$ each period.}
\label{ap}
\end{figure}
\begin{figure}[h]
\includegraphics[scale=0.5]{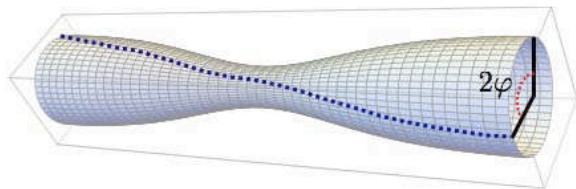}
\caption{Crystalline complex kink (\ref{complex}) plotted as a function of $x$ over one period. The cross-section denotes $\Delta(x)=S(x)-iP(x)$, and indicates a net rotation through the twist parameter $2\varphi$ over one period.}
\label{twist}
\end{figure}

We have found a nontrivial crystalline solution to (\ref{nlse})
\begin{eqnarray}
\Delta(x)&=&A \frac{\sigma(i A\, x+\theta/2-{\bf K})}{\sigma(i A\, x-{\bf K})\sigma(\theta/2)} 
\label{complex}
\\
&& \hskip -1cm \times \,\exp\left[i A\, x \left(-\zeta(\theta/2)+{\rm cs}(\theta/2)\right)+\theta \eta/2+i\,{\rm am}(\theta/2)\right]
\nonumber
\end{eqnarray}
Here, $\sigma$ and $\zeta$ are Weierstrass sigma and zeta functions, $A(\theta)= 2\, {\rm sd}(\theta/4) {\rm cn}(\theta/4)$, and $\eta=\zeta({\bf K})$, with ${\bf K}(\nu)$ the elliptic half-period. For this condensate, both the amplitude and the phase are $x$-dependent, as shown in Figs. \ref{ap}, \ref{twist}.
The essential parameters of the solution (\ref{complex}) are: (i) the parameter $\theta$ which [via (\ref{rotation}) below] characterizes the chiral twist of $\Delta(x)$ over one period; and (ii) the elliptic parameter $\nu$ which, together with $\theta$, determines the crystal period.
The spinor solutions of the Dirac/BdG equation can also be expressed explicitly in terms of elliptic functions, and one can perform the Hartree-Fock analysis, as in the ${\rm GN}_2$ system, to prove self-consistency of the crystalline condensate \cite{bd}. The spectrum is that of a Dirac particle with a {\it single} band in the gap, as shown in Fig. \ref{sheispectrum}. However, unlike the real case where the band lies symmetrically in the center of the gap, here the band is offset. Indeed, the band edges are given by: $E_1=-1$, $E_2=-1+2\, {\rm cn}^2(\theta/4; \nu)$, $E_3=-1+2\, {\rm cd}^2(\theta/4; \nu)$, $E_4=+1$, as shown in Fig. \ref{sheispectrum}. This spectrum is a band version of the Shei spectrum, reducing  to the Shei solution in the infinite period limit. It is also a deformation of the kink crystal spectrum of the [discrete-chiral] ${\rm GN}_2$ system,  reducing to that case when $\theta=2 {\bf K}$. We find the {\it exact} diagonal resolvent (\ref{ansatz}) with
\begin{eqnarray}
{\mathcal N}(E)&=&\frac{1}{4}\frac{1}{\sqrt{(E^2-1)(E-E_2)(E-E_3)}}\quad ,
\end{eqnarray}
$b(E)=2E-(E_2+E_3)$, and 
$a(E)=2(E+1)(E-1-(E_2+E_3)/2)+1+(E_2+E_3)-(E_2-E_3)^2/4$.

Under a shift through one period $L=2{\bf K^\prime}/A$ of the crystal, the BdG Hamiltonian is invariant up to a global chiral rotation through an angle $\varphi$:
\begin{eqnarray}
H\left(x+L\right)&=&e^{i\gamma^5 \varphi}\, H\left(x\right)\, e^{-i\gamma^5 \varphi}\quad ,\label{rotation}
\\
\varphi&=& {\bf K^\prime}\left[-\zeta(\theta/2)+{\rm cs}(\theta/2)-i \theta\, \zeta(i{\bf K^\prime})/(2{\bf K^\prime})\right]\nonumber
\end{eqnarray}
The solutions to $H\psi=E\psi$ acquire a chiral rotation and a Bloch phase under a period shift, $\psi(x+L)=e^{i k L} e^{i\varphi\gamma^5} \psi(x)$. The [relativistic] Bloch momentum $k$ is related to the the spectral function by $\rho(E)=dk/dE$.

\begin{figure}[h]
\includegraphics[scale=0.35]{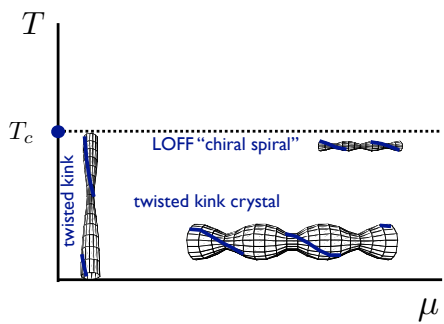}
\caption{Phase diagram of the chiral ${\rm NJL}_2$ model, from a Ginzburg-Landau analysis based on a crystalline condensate of the form in (\ref{complex}). Along the $T$ axis (below $T_c$) the condensate is of the form of Shei's twisted kink (\ref{shei}), while on the $T=T_c$ line the condensate reduces to the form of the LOFF chiral spiral \cite{thies-njl}. The condensate is depicted as in Fig. \ref{twist}.}
\label{phase}
\end{figure}

To conclude, we discuss briefly the implications of this self-consistent solution to the gap equation for the 
$(T,\mu)$ phase diagram of the ${\rm NJL}_2$ model. Recall that the corresponding real condensate characterizes the inhomogeneous crystalline phase of the ${\rm GN}_2$ system \cite{thies-gn}.  For the ${\rm NJL}_2$ model, the Ginzburg-Landau (GL) approach shows that the "chiral spiral" phase identified in \cite{thies-njl} has a richer structure, characterized by a crystalline complex condensate of the form (\ref{complex}).
Near the tricritical point of the  massless ${\rm NJL}_2$ system the  GL effective Lagrangian is 
\begin{eqnarray}
{\mathcal L}_{GL}&=&c_0+c_2|\Delta|^2 +c_3 {\rm Im}\left[\Delta (\Delta^\prime)^*\right] +c_4\left[|\Delta|^4+|\Delta^\prime |^2\right]\nonumber\\
&&+c_5{\rm Im}\left[ \left(\Delta^{\prime\prime}-3|\Delta |^2\Delta\right)(\Delta^\prime)^*\right]  \\
&&\hskip -1cm 
+c_6\left[ 2|\Delta |^6+8 |\Delta |^2 |\Delta^\prime |^2+2 {\rm Re}\left((\Delta^\prime)^2(\Delta^*)^2\right)
+|\Delta^{\prime\prime}|^2\right]\nonumber
\label{gl}
\end{eqnarray}
Here the coefficients $c_n$ are known functions of $T$ and  $\mu$ \cite{buzdin3,thies-gl}. (This GL approach has been used previously in \cite{thies-gl} to describe the phase diagram of the massive and massless ${\rm NJL}_2$ models, in the vicinity of the tricritical point. For the {\it massive} ${\rm NJL}_2$ model no (complex) exact solution to the gap equation is known, so \cite{thies-gl} is the current state-of-the-art for the massive model.) In the ${\rm GN}_2$ model, which has a real condensate, there is a tricritical point 
at $T=0.3183$, $\mu=0.6082$; given by the point  $c_2=c_4=0$. In the [chiral] ${\rm NJL}_2$ model, which has a complex condensate, there is a tricritical point at $T=0.5669$, $\mu=0$; given by  $c_2=c_3=0$. To search for possible crystalline phases near the tricritical point we keep terms up to $c_4$, and study the effective equation of motion
\begin{equation}
c_4 \Delta^{\prime\prime}-i c_3\Delta^\prime-\left(c_2+2 c_4 |\Delta|^2\right)\Delta =0 \quad .
\label{gleq}
\end{equation}
Note that this has precisely the same form as the NLSE (\ref{nlse}) found from the gap equation and the Eilenberger equation. Thus, we can use our solution (\ref{complex}) as a variational ansatz and compute the free energy. We have found that the resulting free energy is lower than that of the LOFF-form ``chiral spiral'' variational ansatz $\Delta(x)=A e^{iqx}$, for $T<T_c$. Noting that our ansatz reduces to the chiral spiral in the perturbative limit (just as the ${\rm GN}_2$ self-consistent crystal condensate reduces to the LOFF form $\Delta(x)=A \sin(qx)$ in this limit \cite{thies-gn}), we are led to the phase diagram shown in Fig. \ref{phase}. For a detailed discussion of the phase diagram, based on the {\it full} free energy (beyond the GL approximation), using our exact spectral data, see \cite{bd}.



\begin{acknowledgments}
We thank the DOE for support through grant DE-FG02-92ER40716.
\end{acknowledgments}


\begin{thebibliography}{99}

\bibitem{campbell}
  D.~K.~Campbell and A.~R.~Bishop,
  ``Soliton Excitations In Polyacetylene And Relativistic Field Theory Models,''
  Nucl.\ Phys.\  B {\bf 200}, 297 (1982).

  \bibitem{rajagopal}
  K.~Rajagopal and F.~Wilczek,
  ``The condensed matter physics of QCD,'' 
  in {\it At the Frontier of Particle Physics / Handbook of QCD}, M. Shifman, ed., (World Scientific, 2001);
  arXiv:hep-ph/0011333.
  
   \bibitem{casalbuoni}
  R.~Casalbuoni and G.~Nardulli,
  ``Inhomogeneous superconductivity in condensed matter and QCD,''
  Rev.\ Mod.\ Phys.\  {\bf 76}, 263 (2004).
  [arXiv:hep-ph/0305069].
  
  \bibitem{pitaevskii}
S.~Giorgini, L.~P.~Pitaevskii and S.~Stringari, 
``Theory of ultracold Fermi gases'', 
[arXiv:0706.3360], to appear in Rev.\ Mod.\ Phys.
  
  \bibitem{peierls}
R.~Peierls, {\it The Quantum Theory of Solids} (Oxford, 1955).

\bibitem{degennes}
P.~G.~de~Gennes, {\it Superconductivity of Metals and Alloys} (Addison-Wesley, Redwood City, CA, 1989).

\bibitem{nambu}
  Y.~Nambu and G.~Jona-Lasinio,
  ``Dynamical model of elementary particles based on an analogy with superconductivity,''
  Phys.\ Rev.\  {\bf 122}, 345 (1961).

\bibitem{gross}
  D.~J.~Gross and A.~Neveu,
  ``Dynamical Symmetry Breaking In Asymptotically Free Field Theories,''
  Phys.\ Rev.\  D {\bf 10}, 3235 (1974).
  
  \bibitem{dhn}
  R.~F.~Dashen, B.~Hasslacher and A.~Neveu,
  ``Semiclassical Bound States In An Asymptotically Free Theory,''
  Phys.\ Rev.\  D {\bf 12}, 2443 (1975).
  
 \bibitem{shei}
  S.~S.~Shei,
  ``Semiclassical Bound States In A Model With Chiral Symmetry,''
  Phys.\ Rev.\  D {\bf 14}, 535 (1976).
  
  \bibitem{feinberg}
  J.~Feinberg and A.~Zee,
  ``Dynamical Generation of Extended Objects in a $1+1$ Dimensional Chiral
 Field Theory: Non-Perturbative Dirac Operator Resolvent Analysis,''
  Phys.\ Rev.\  D {\bf 56}, 5050 (1997)
  [arXiv:cond-mat/9603173];
  ``Generalized supersymmetric quantum mechanics and reflectionless fermion
  bags in 1+1 dimensions,'' 
  in {\it Multiple facets of quantization and supersymmetry}, M.~Olshanetsky {\it et al} (ed.) (World Scientific, 2002) 
  arXiv:hep-th/0109045.
  
 \bibitem{thies-njl}
  V.~Schon and M.~Thies,
  ``Emergence of Skyrme crystal in Gross-Neveu and 't Hooft models at  finite density,''
  Phys.\ Rev.\  D {\bf 62}, 096002 (2000).
  [arXiv:hep-th/0003195].
  
  \bibitem{thies-gn}
M.~Thies,
``Analytical solution of the Gross-Neveu model at finite density,''
  Phys.\ Rev.\  D {\bf 69}, 067703 (2004),
  [arXiv:hep-th/0308164];
  ``From relativistic quantum fields to condensed matter and back again:
   Updating the Gross-Neveu phase diagram,''
  J.\ Phys.\ A  {\bf 39}, 12707 (2006).
  [arXiv:hep-th/0601049].
  
  \bibitem{deforcrand}
  P.~de Forcrand and U.~Wenger,
  ``New baryon matter in the lattice Gross-Neveu model,''
  PoS {\bf LAT2006}, 152 (2006).
  [arXiv:hep-lat/0610117].
  
  \bibitem{wolff}
U.~Wolff, 
``The phase diagram of the infinite-N Gross-Neveu Model at finite temperature and chemical potential'',
 Phys.\ Lett.\ {\bf 157B}, 303 (1985).

\bibitem{horovitz}
B.~Horovitz,
``Soliton Lattice in Polyacetylene, Spin-Peierls Systems, and Two-Dimensional Sine-Gordon 
 Systems'',
Phys.\ Rev.\ Lett.\ {\bf 46}, 742  (1981).

\bibitem{braz}
S.~A.~Brazovskii, S.~A.~Gordynin, and N.~N.~Kirova,
``Exact solution of the Peierls model with an arbitrary number of electrons in the unit cell'',
Pis. Zh. Eksp. Teor. Fiz. {\bf 31}, 486 (1980) [JETP Lett. {\bf 31}, 456 (1980)];
S.~A.~Brazovskii and N.~N.~Kirova, 
``Excitons, polarons and bipolarons in conducting polymers'', Pis. Zh. Eksp. Teor. Fiz. {\bf 33}, 6 (1981) 
[JETP Lett. {\bf 33}, 4 (1981)].

\bibitem{mertsching}
J.~Mertsching and H.~J.~Fischbeck,
``The incommensurate Peierls phase of the quasi-dimensional Frohlich model with a nearly half-filled band'', 
Phys. Stat. Sol. B {\bf 103}, 783 (1981).

\bibitem{buzdin1}
A.~I.~Buzdin and V.~V.~Tugushev, 
``Phase diagrams of electronic and superconducting transitions to soliton lattice states'',
Zh. Eksp. Teor. Fiz. {\bf 85}, 735 (1983), [Sov. Phys. JETP {\bf 58}, 428 (1983).

\bibitem{machida}
K.~Machida and H.~Nakanishi,
``Superconductivity under a ferromagnetic molecular field'', 
Phys.\ Rev.\ B {\bf 30}, 122  (1984).

\bibitem{jackiw}
  R.~Jackiw and C.~Rebbi,
  ``Solitons With Fermion Number 1/2,''
  Phys.\ Rev.\  D {\bf 13}, 3398 (1976).
  
  \bibitem{gw}
  J.~Goldstone and F.~Wilczek,
  ``Fractional Quantum Numbers On Solitons,''
  Phys.\ Rev.\ Lett.\  {\bf 47}, 986 (1981).


\bibitem{zwierlein} M.~W.~Zwierlein, A.~Schirotzek, C.~H.~Schunck and W.~Ketterle, ``Fermionic Superfluidity with Imbalanced Spin Populations and the Quantum Phase Transition to the Normal State'',
Science {\bf 311}, 492 (2006).

\bibitem{partridge} G.~B.~Partridge {\it et al}, 
 ``Pairing and Phase Separation in a Polarized Fermi Gas'', 
Science {\bf 311}, 503 (2006).

  


\bibitem{eilenberger}
G.~Eilenberger, 
``Transformation of Gorkov's Equation for Type II Superconductors into Transport Like Equations'', 
Z. Phys. {\bf 214}, 195 (1968).










\bibitem{sakita}
  Z.~b.~Su and B.~Sakita,
  ``Chiral symmetry and chiral anomaly in an incommensurate charge-density-wave system,''
  Phys.\ Rev.\ Lett.\  {\bf 56}, 780 (1986).

\bibitem{stone}
I.~Kosztin, S.~Kos, M.~Stone and A.~J.~Leggett, 
``Free energy of an inhomogeneous superconductor: A wave-function approach'',
Phys.\ Rev.\ B {\bf 58}, 9365 (1998); 
S.~Kos and  M.~Stone, 
``Gradient expansion for the free energy of a clean superconductor '', 
Phys.\ Rev.\ B {\bf 59}, 9545 (1999).

\bibitem{dikii}
L.~A.~Dickey, {\it Soliton Equations and Hamiltonian Systems},  (World Scientific, Singapore, 1991).

\bibitem{braz2} 
S.~A.~Brazovskii, N.~N.~Kirova and S.~I.~Matveenko,
``Peierls effect in conducting polymers'', 
Zh. Eksp. Teor. Fiz. {\bf 86}, 743 (1984)
Sov.\ Phys.\ JETP {\bf 59}, 434 (1984).


\bibitem{bd}
  G.~Basar and G.~V.~Dunne, in preparation.

\bibitem{karbstein}
  F.~Karbstein and M.~Thies,
  ``Divergence of the axial current and fermion density in Gross-Neveu models,''
  Phys.\ Rev.\  D {\bf 76}, 085009 (2007).
  [arXiv:0706.0424 [hep-th]].
  
\bibitem{buzdin3}
A.~I.~Buzdin and H.~Kachkachi,  ``Generalized Ginzburg-Landau theory for nonuniform FFLO superconductors'', Phys.\ Lett.\ A\ {\bf 225}, 341 (1997).

\bibitem{thies-gl}
C.~Boehmer, M.~Thies and K.~Urlichs,
 ``Tricritical behavior of the massive chiral Gross-Neveu model,''
 Phys.\ Rev.\  D {\bf 75}, 105017 (2007).





 























%


%

%

  


\end{thebibliography}

\end{document}